# Experimental setup and procedure for the measurement of the $^7$Be(n,$\alpha$)$\alpha$ reaction at n_TOF

L. Cosentino[1], A. Musumarra[1,2], M. Barbagallo[3], A. Pappalardo[1], N. Colonna[3], L. Damone[3], M. Piscopo[1], P. Finocchiaro[1,*], E. Maugeri[4], S. Heinitz[4], D. Schumann[4], R. Dressler[4], N. Kivel[4], O. Aberle[5],
J. Andrzejewski[6], L. Audouin[7], M. Ayranov[8], M. Bacak[9], S. Barros[10], J. Balibrea-Correa[11], V. Bécares[11], F. Bečvář[12], C. Beinrucker[13], E. Berthoumieux[14], J. Billowes[15], D. Bosnar[16], M. Brugger[5], M. Caamaño[17], M. Calviani[5], F. Calviño[18], D. Cano-Ott[11], R. Cardella[1,5], A. Casanovas[18], D. M. Castelluccio[19,20], F. Cerutti[5], Y. H. Chen[7], E. Chiaveri[5], G. Cortés[18], M. A. Cortés-Giraldo[21], M. Diakaki[14], C. Domingo-Pardo[22], E. Dupont[14], I. Duran[17], B. Fernandez-Dominguez[17], A. Ferrari[5], P. Ferreira[10], W. Furman[23], S. Ganesan[24], A. García-Rios[11], A. Gawlik[6], I. Gheorghe[25], T. Glodariu[25], K. Göbel[13], I. F. Gonçalves[10], E. González-Romero[11], E. Griesmayer[9], C. Guerrero[21], F. Gunsing[14], H. Harada[26], T. Heftrich[13], J. Heyse[27], D. G. Jenkins[28], E. Jericha[9], F. Käppeler[29], T. Katabuchi[30], P. Kavrigin[9], A. Kimura[26], M. Kokkoris[31], M. Krtička[12], E. Leal-Chidonca[17], J. Lerendegui[21], C. Lederer[32], H. Leeb[9], S. Lo Meo[19,20], S. Lonsdale[32], R. Losito[5], D. Macina[5], J. Marganiec[6], T. Martínez[11], C. Massimi[33,20], P. Mastinu[34], M. Mastromarco[3], F. Matteucci[35], A. Mazzone[36], E. Mendoza[11], A. Mengoni[19], P. M. Milazzo[35], F. Mingrone[20], M. Mirea[25], S. Montesano[5], R. Nolte[37], A. Oprea[25], N. Patronis[38], A. Pavlik[39], J. Perkowski[6], J. Praena[21,40], J. Quesada[21], K. Rajeev[24], T. Rauscher[41,42], R. Reifarth[13], A. Riego-Perez[18], P. Rout[24], C. Rubbia[5], J. Ryan[15], M. Sabate-Gilarte[5], A. Saxena[24], P. Schillebeeckx[27], S. Schmidt[13], P. Sedyshev[23], A. Stamatopoulos[31], G. Tagliente[3], J. L. Tain[22], A. Tarifeño-Saldivia[22], L. Tassan-Got[7], A. Tsinganis[31], S. Valenta[12], G. Vannini[33,20], V. Variale[3], P. Vaz[10], A. Ventura[20], V. Vlachoudis[5], R. Vlastou[31], J. Vollaire[5], A. Wallner[43,39], S. Warren[15], M. Weigand[13], C. Weiß[5], C. Wolf[13], P. J. Woods[32], T. Wright[15], P. Žugec[16]

(n_TOF Collaboration)[#]

1. *INFN Laboratori Nazionali del Sud, Catania, Italy*
2. *Dipartimento di Fisica e Astronomia DFA, Università di Catania, Italy*
3. *INFN Sezione di Bari, Italy*
4. *Paul Scherrer Institut, Nuclear Energy and Safety Research Department, 5232 Villigen, Switzerland*
5. *CERN, Geneva, Switzerland*
6. *Uniwersytet Łódzki, Lodz, Poland*
7. *Institut de Physique Nucléaire, CNRS-IN2P3, Univ. Paris-Sud, Université Paris-Saclay, 91406 Orsay Cedex, France*
8. *European Commission, DG-Energy, Luxembourg*
9. *Atominstitut der Österreichischen Universitäten, Technische Universität Wien, Austria*
10. *C2TN-Instituto Superior Tecníco, Universidade de Lisboa, Portugal*
11. *Centro de Investigaciones Energeticas Medioambientales y Tecnológicas (CIEMAT), Madrid, Spain*
12. *Charles University, Prague, Czech Republic*
13. *Johann-Wolfgang-Goethe Universität, Frankfurt, Germany*
14. *CEA/Saclay-IRFU, Gif-sur-Yvette, France*
15. *University of Manchester, Oxford Road, Manchester, United Kingdom*
16. *Department of Physics, Faculty of Science, University of Zagreb, Croatia*
17. *Universidade de Santiago de Compostela, Galicia, Spain*
18. *Universitat Politecnica de Catalunya, Barcelona, Spain*
19. *ENEA, Bologna, Italy*

20. *INFN Sezione di Bologna, Italy*
21. *Universidad de Sevilla, Seville, Spain*
22. *Instituto de Física Corpuscular, CSIC-Universidad de Valencia, Spain*
23. *Joint Institute of Nuclear Research, Dubna, Russia*
24. *Bhabha Atomic Research Centre (BARC), Mumbai, India*
25. *Horia Hulubei National Institute of Physics and Nuclear Engineering-IFIN HH, Bucharest-Magurele, Romania*
26. *Japan Atomic Energy Agency (JAEA), Tokai-mura, Japan*
27. *European Commission JRC, Institute for Reference Materials and Measurements, Retieseweg 111, B-2440 Geel, Belgium*
28. *University of York, Heslington, York, United Kingdom*
29. *Karlsruhe Institute of Technology (KIT), Institut für Kernphysik, Karlsruhe, Germany*
30. *Tokyo Institute of Technology, Japan*
31. *National Technical University of Athens (NTUA), Athens, Greece*
32. *University of Edinburgh, UK*
33. *Dipartimento di Fisica, Università di Bologna, Italy*
34. *INFN Laboratori Nazionali di Legnaro, Italy*
35. *INFN Sezione di Trieste, Italy*
36. *Institute of Crystallography, CNR, Bari, Italy*
37. *Physikalisch Technische Bundesanstalt (PTB), Braunschweig*
38. *University of Ioannina, Greece*
39. *University of Vienna, Faculty of Physics, Vienna, Austria*
40. *University of Granada, Spain*
41. *Centre for Astrophysics Research, School of Physics, Astronomy and Mathematics, University of Hertfordshire, Hatfield, United Kingdom*
42. *Department of Physics, University of Basel, Basel, Switzerland*
43. *Research School of Physics and Engineering, Australian National University, Canberra ACT 0200, Australia*

* Corresponding author FINOCCHIARO@LNS.INFN.IT
# www.cern.ch/ntof

**Abstract**

The newly built second experimental area EAR2 of the n_TOF spallation neutron source at CERN allows to perform (n, charged particles) experiments on short-lived highly radioactive targets. This paper describes a detection apparatus and the experimental procedure for the determination of the cross-section of the $^{7}$Be(n,$\alpha$) reaction, which represents one of the focal points toward the solution of the cosmological Lithium abundance problem, and whose only measurement, at thermal energy, dates back to 1963.

The apparently unsurmountable experimental difficulties stemming from the huge $^{7}$Be $\gamma$-activity, along with the lack of a suitable neutron beam facility, had so far prevented further measurements. The detection system is subject to considerable radiation damage, but is capable of disentangling the rare reaction signals from the very high background. This newly developed setup could likely be useful also to study other challenging reactions requiring the detectors to be installed directly in the neutron beam.

## 1 Introduction

One of the current challenges in nuclear physics is related to the measurement of neutron-induced reactions on short-lived radioisotopes, which often exhibit very low cross-sections associated with high specific activity. The newly built second experimental area EAR2 of the

n_TOF neutron-time-of-flight facility at CERN allows to perform (n, charged particles) experiments on short-lived highly radioactive targets, as low-mass samples can be used thanks to the high neutron flux [1],[2],[3]. This paper describes the experimental setup developed in order to perform one of these challenging measurements, namely the neutron capture reaction of $^{7}$Be, which is important for the so-called "Cosmological Lithium problem" (CLIP), one of the most relevant problems in astrophysics nowadays [4],[5]. Among the key nuclear reactions, whose cross-section could shed some light on the CLIP, the $^{7}$Be(n,$\alpha$)$\alpha$ cross-section has been awaited since more than fifty years. The only measurement in literature for this reaction channel was performed in 1963 by using thermal neutrons, thus very far from the relevant astrophysical energy range [6]. The $^{7}$Be(n,$\alpha$)$\alpha$ reaction considered in this work has two alpha particles in the final state, emitted back to back with a total kinetic energy of 16.6 MeV (there are other $^{8}$Be levels that can decay through the two-alpha channel, in particular at 3 and 18.9 MeV, but in the first case the alpha-particles are emitted with a kinetic energy too small to be detected, while in the second case the cross section is extremely low [6],[7]). Detecting both alpha particles with $\approx$ 8 MeV energy allows for an efficient background suppression. As a consequence an isotopically pure $^{7}$Be target is not strictly needed due to the uniqueness of the decay signature. A major difficulty in performing the experiment lies in the extremely high specific activity of the $^{7}$Be target (13 GBq/μg), which decays by electron capture to the ground state of $^{7}$Li (Branching Ratio $\approx$ 90%) or to the first excited state followed by emission of a 478 keV γ-ray. Apart from representing a serious radioprotection issue, this γ-ray load strongly affects the detection system that has also to withstand the neutron beam. On top of this, the huge γ-background is deteriorating the signal-to-background ratio, given the small reaction cross-section expected in the range of hundreds of millibarn. Additional difficulties are the background and the possible detector damage due to the intense neutron beam accompanied by the so-called γ-flash, the large prompt flux of γ-rays and relativistic charged particles produced by the n_TOF spallation target.

The improved neutron-flux performances of the EAR2 facility, and the perfectly matched n_TOF dynamic range, offer the opportunity to measure this cross-section in a wide energy range for the first time. The advantage of EAR2, with respect to other neutron facilities in the world, is the extremely high instantaneous neutron flux ($10^{7}\div10^{8}$ neutrons/pulse) delivered in a short time interval (about 10 ms) at the sample position, which results in a favorable signal-to-background ratio when measuring on small mass and/or highly radioactive targets, as well as on isotopes characterized by a small reaction cross-section.

In the following sections the detection technique, detectors and mechanical structure are described, along with the challenging requirements and the preliminary tests aimed at proving the feasibility of the experiment. The detailed physics results of the measurement are going to be published soon [8].

## 2 The detectors

The specific detection setup design was conceived following three main directives:

- *large geometric efficiency*;
- *redundancy,* by doubling the target-detection setup for a fast and reliable characterization of systematic errors;
- *$\alpha-\alpha$ coincidence detection mode,* for suppression of neutron- and γ-induced backgrounds.

This approach led to a setup based on single pad Silicon detectors. The most demanding tasks for the apparatus were: (i) withstanding the radiation damage without major degradation in detection performance; (ii) picking out the alpha signals from the huge background.

Silicon detectors were chosen because of their absolute energy response and low sensitivity to γ-rays and neutrons. Indeed, low energy γ-rays and neutrons up to $\approx$ 1 MeV can only produce low amplitude signals in Silicon, with an interaction probability $< 10^{-3}\div10^{-4}$. For more energetic neutrons and γ-rays the probability of higher energy deposition goes down to the order of $10^{-5} \div 10^{-6}$ [9]. The basic detector unit was a square pad, 3cm x 3cm active area and 140 μm thick, produced by Micron Semiconductor Ltd [10]. Four such detectors were assembled in form of two independent sandwiches, each one ready to host one $^{7}$Be target. The distance between the two targets was 30 mm, the distance between a target and the adjacent detectors was 7 mm (Figure 1). The sandwich configuration resulted in a large solid angle, thus maximizing the detection efficiency and optimizing the usage of the neutron beam. Moreover, as the final state of the reaction under study consists of two ≈ 8 MeV alpha particles emitted back-to-back, the huge background can be strongly suppressed by suitable energy thresholds and a narrow time coincidence window. Furthermore, by analyzing the coincidences between detectors from different sandwiches, one can estimate the rate of random spurious events, thus optimizing the choice of the best threshold values for the offline analysis. It has to be remarked that the range of the expected alpha particles in Silicon is below 60 μm, thus preventing cross-talk between the sandwiches. Finally, the use of two independent setups, with targets produced by different methods, allows to cross-check the results by verifying that they lead to the same physical outcome. The differences, if any, provide a hint of the systematic errors, and in the end one can sum the two data sets, thus improving the statistics.

The energy resolution of the Silicon detectors of about 1% was measured with an $^{241}$Am source, a value that exceeds the needs for the discrimination of the high energy alphas from the background. However, as the expected fluence was $\approx 10^{12} \div 10^{13}$ neutrons, a clear degradation was expected during the measurement, although still preserving a good discrimination capability.

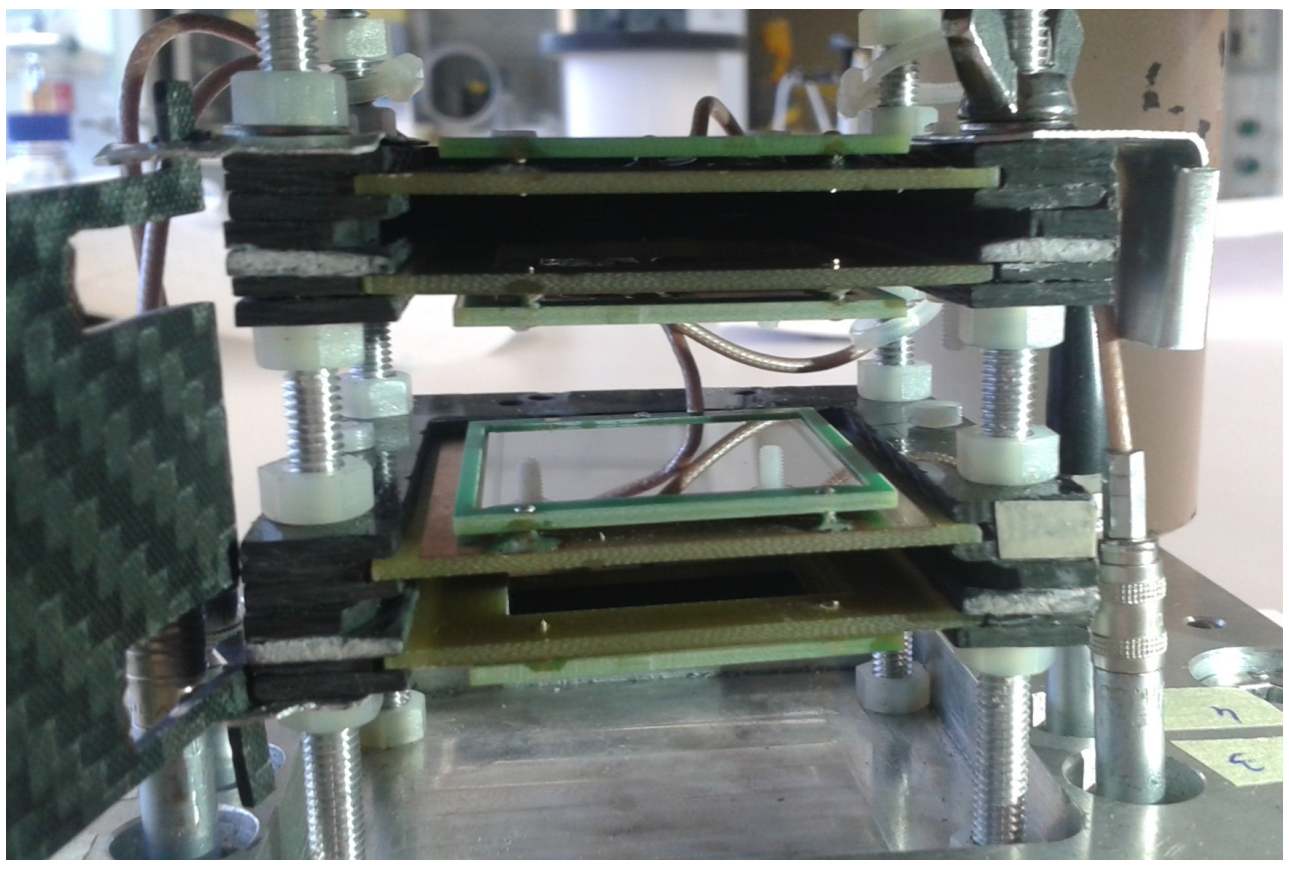

Figure 1. The carbon fibre structure to host the two Silicon-target-Silicon sandwiches. The four Silicon detectors are already assembled (the surface of the top detector of the lower sandwich is visible), whereas the two $^{7}$Be targets were mounted later by means of a robotic manipulator in a hot cell.

The detection efficiency of each sandwich was simulated by means of a Monte Carlo code, taking into account the known XY profile of the neutron beam (Figure 2) as well as the spatial profile of the $^{7}$Be deposit. As the detectors do not cover the full 4π solid angle, the single α detection efficiency resulted 67%, whereas the α-α coincidence efficiency is 39.7% for one target and 36.9% for the other one. The reason of such a low coincidence detection efficiency, in spite of the almost full coverage of the sandwich configuration, is that the reactions are induced in the $^{7}$Be sample almost uniformly over the whole surface, which matches the 3x3 cm$^{2}$ of the detectors, so that there is a considerable loss of efficiency for the events occurring near the borders. For reactions occurring in the peripheral region of the target at least one alpha will almost certainly be detected whereas the other one is likely to be lost.

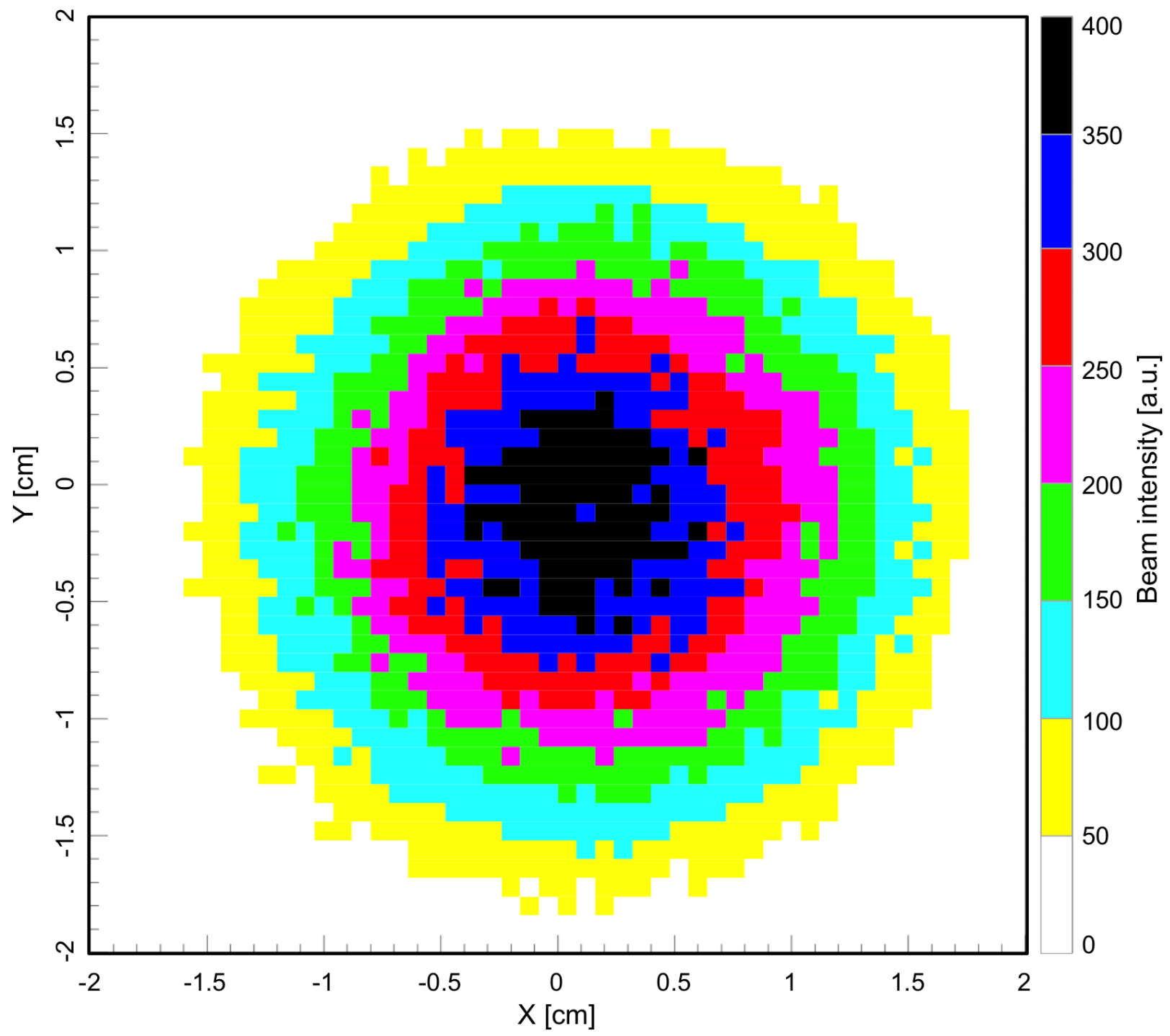

Figure 2. The known XY neutron beam profile at the measurement position.

## 3 The validation tests

In order to assess the feasibility of the measurement of the $^{7}$Be(n,α) cross section at n_TOF, the following four main issues had to be addressed:

- Q1. Would the detectors survive in the intense neutron flux of EAR2?
- Q2. Would the resolution degradation still allow to discriminate the alpha particles from the background?
- Q3. Could an electronic setup be fast enough to recover after the γ-flash, in time to register events produced by neutrons of energy up to at least 10 keV?
- Q4. Would the detectors and the front-end electronics be capable of detecting and discriminating the alpha particle signals when exposed to the very intense γ-ray flux from the $^{7}$Be targets?

Therefore, the operational capability and radiation hardness were tested by a long irradiation of two detectors with the neutron beam at EAR2, followed by a test with a low activity $^{7}$Be sample at PSI, and finally by a test with a 39 GBq $^{137}$Cs γ-source at INFN-LNS.

### *3.1 Test with neutron beam and $^{6}$LiF target*

To answer question Q1, a test prototype sandwich was built and mounted in the EAR2 beam line close to the beam dump, that is in an extremely harsh environment. Between the two detectors a neutron converter was inserted, namely a 105 μg/cm$^{2}$ layer of $^{6}$LiF deposited onto a 1.5 μm thick Mylar foil (Figure 3). $^{6}$Li is a well-known neutron converter, by means of the reaction

$$^{6}Li + n \rightarrow {}^{3}H\ (2.73\ MeV) + \alpha\ (2.05\ MeV) \quad (1)$$

which has a 940 b cross section at thermal neutron energy that decreases with the usual 1/v behavior. The two particles are emitted back to back in the center of mass reference frame.

This setup was exposed to the neutron beam with a total fluence integrated over the detector surface of more than $10^{12}$ neutrons, i.e. about half the foreseen fluence for the real experiment. The reverse current was periodically checked throughout this period, and it increased as expected due to the progressive radiation damage.

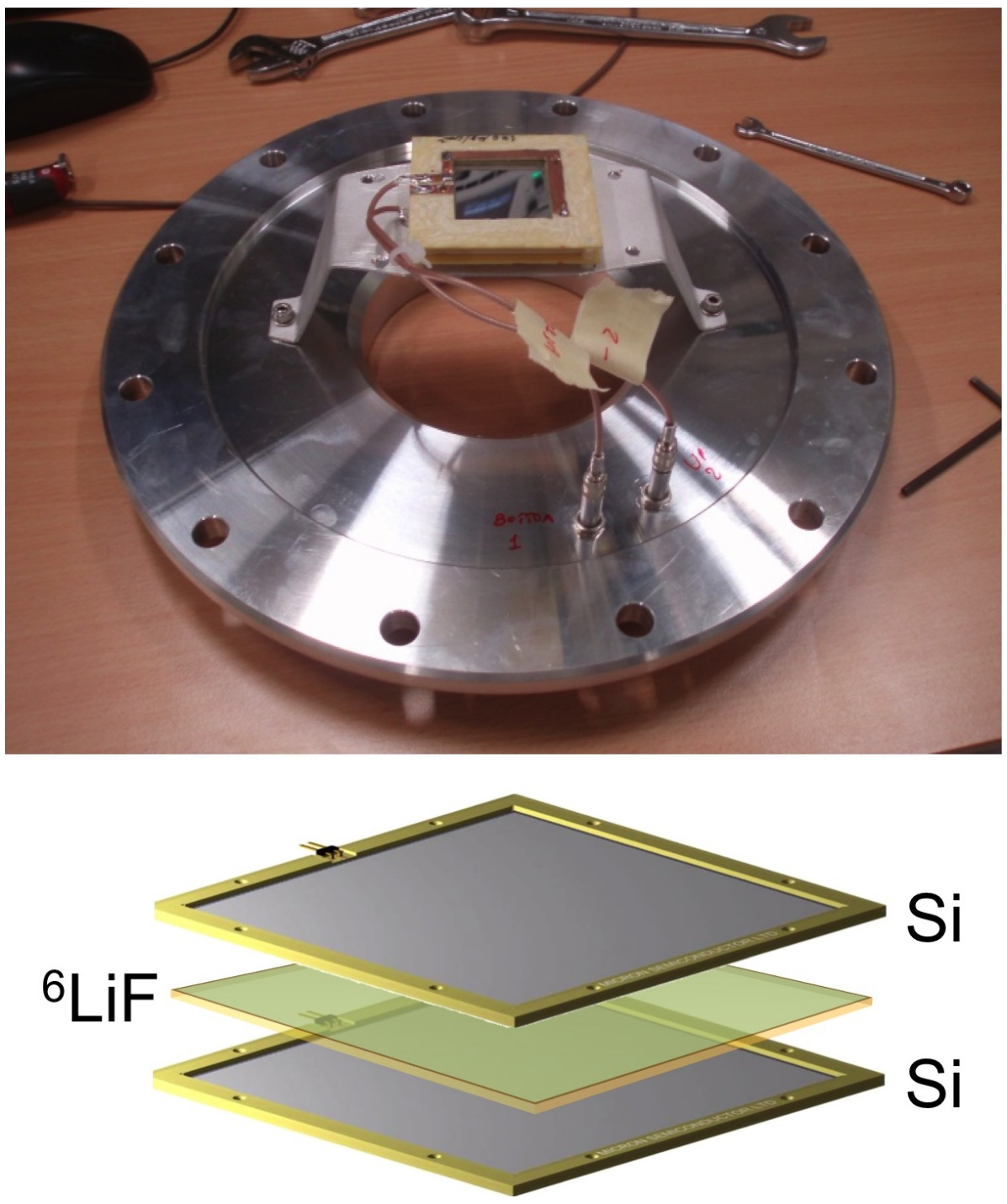

Figure 3. The Silicon-$^{6}$LiF-Silicon sandwich employed for the feasibility test in the EAR2 neutron beam dump.

The increase in reverse current implies a decrease of the resistivity, and causes a decrease of the effective voltage bias on the detector. Therefore, the operating voltage had to be correspondingly increased periodically, in order to restore a correct bias on the detectors. By the end of the irradiation period a reverse current of about 300 nA was reached (Figure 4), well below the operational limit of several microamps and in agreement with what has been observed on the SIMON2D Silicon monitor after scaling for the different thickness [11]. As the expected duration of the $^{7}$Be(n,$\alpha$) measurement was about 45 days with a total fluence (integrated over the sample area) of $\approx 3 \cdot 10^{12}$ neutrons, the test showed that the resistance of the detectors to radiation damage was acceptable.

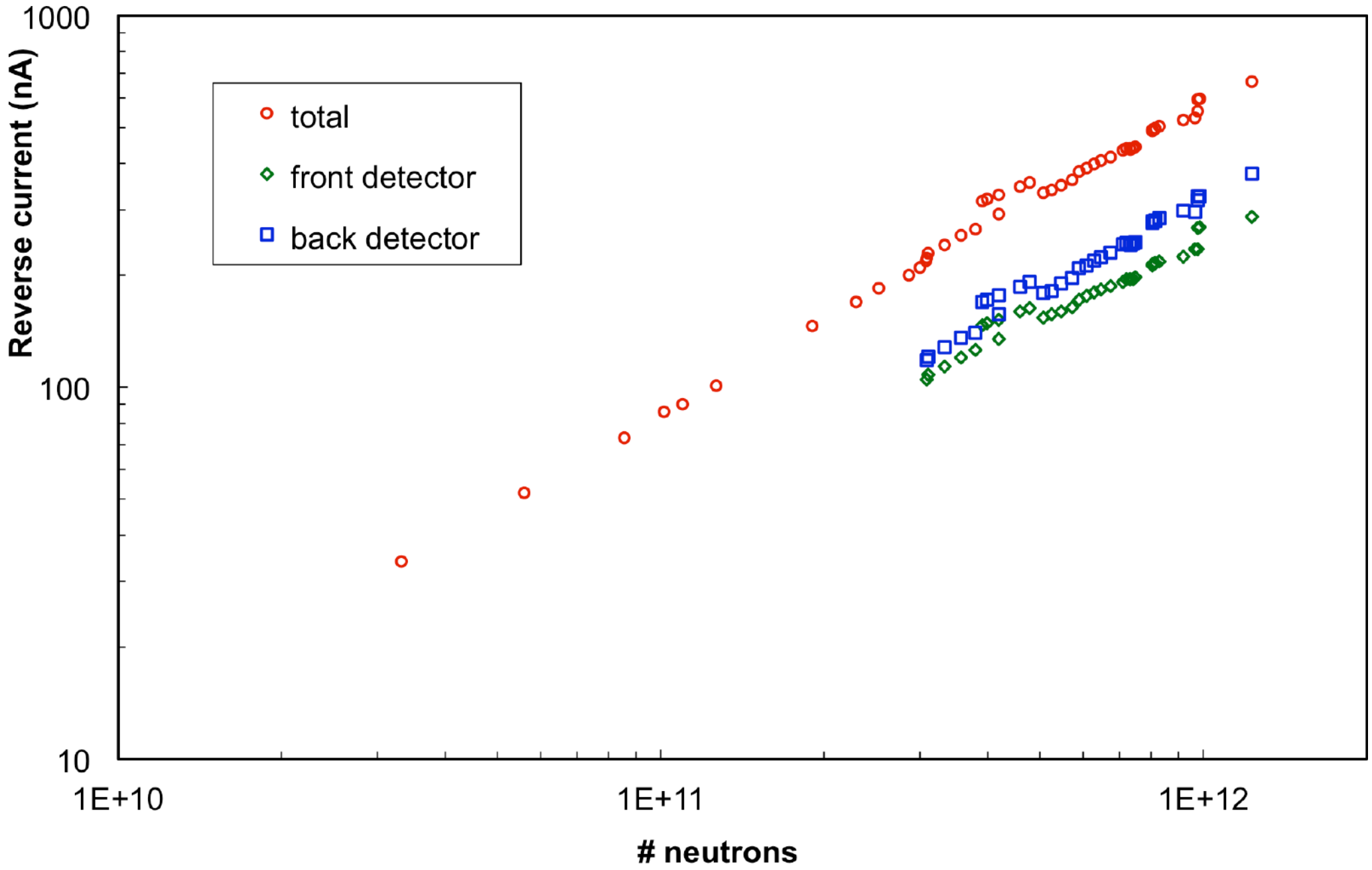

Figure 4. The measured reverse current as a function of the neutron fluence for the two test detectors. Also reported is the sum of the two currents.

According to question Q2 the resolution of the setup under the expected irradiation conditions has been checked using the separation of the 2.73 MeV tritons from the 2.05 MeV alpha particles. If the two groups could still be resolved after the test irradiation, one could reasonably expect that by the end of the effective measurement the 8 MeV alphas emitted in the $^{7}$Be(n,α) reaction could still be separated from any background events.

The energy spectra measured by both detectors at the beginning of the irradiation are shown in Figure 5. The front detector was the one directly facing the $^{6}$LiF layer, the back one faced the Mylar substrate. The threshold was set around 1.5 MeV to suppress the huge background originating from the nearby beam dump. One can immediately see that the triton peak in the back detector is slightly shifted towards lower energy, as expected due to the presence of the Mylar. For the same reason the alpha peak is almost completely shifted below the threshold. In Figure 6 the spectrum of the back detector is shown, as measured at the beginning and shortly before the end of the irradiation test. Even though a degradation of the energy resolution due to neutron irradiation is visible, it does not prevent the selection of the triton peak.

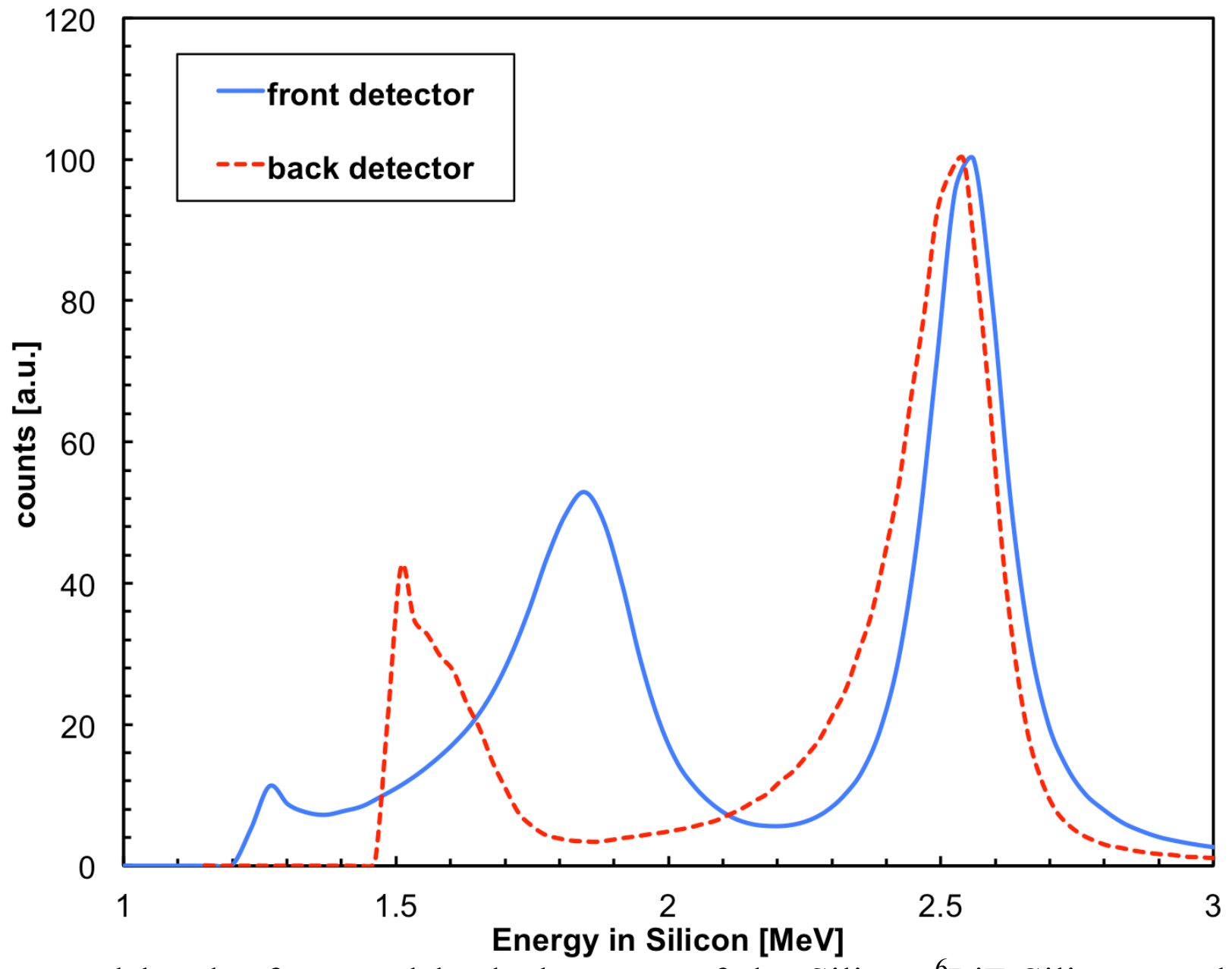

Figure 5. Spectra measured by the front and back detectors of the Silicon-$^6$LiF-Silicon sandwich employed for the feasibility test in the EAR2 neutron beam. Both alpha and triton peaks are clearly visible in the front detector spectrum (solid line). The spectrum of the back detector is shifted towards lower energies due to the effect of the Mylar backing, mostly affecting the alpha peak which is partly cut off by the electronic threshold.

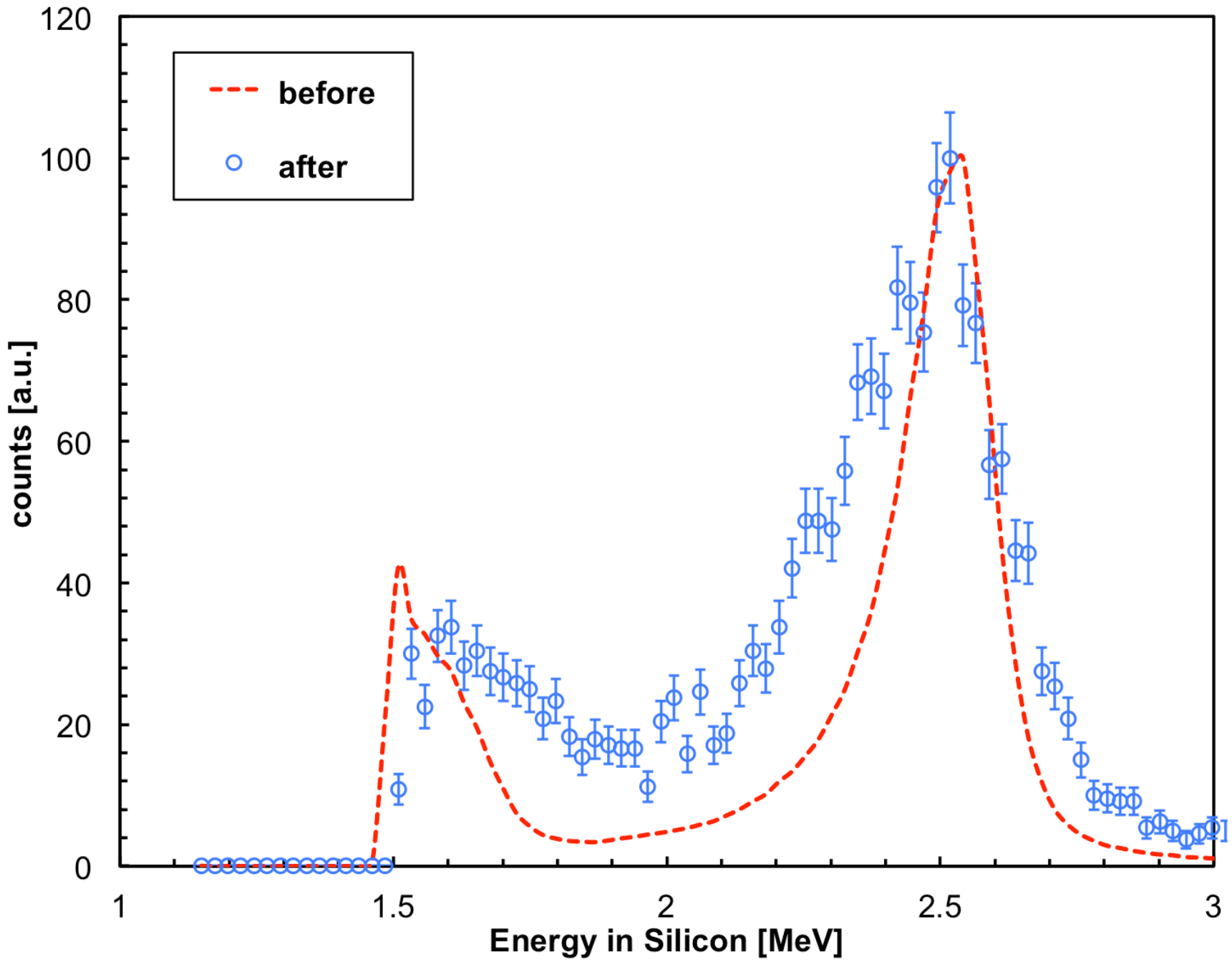

Figure 6. Spectrum of the back detector of the Silicon-$^6$LiF-Silicon sandwich, measured at the beginning (dashed line) and shortly at the end of the irradiation test (open circles). Even though a degradation of the energy resolution due to neutron irradiation is clearly visible, the selection of the triton peak is still possible.

Q3 imposed the need for fast preamplifiers capable of withstanding the γ-flash. To identify time-of-flight signals corresponding to several keV neutron energy, the preamplifiers had to recover from the γ-flash as fast as possible, in any case within 20 μs. A few experimental observations of the waveforms during the test irradiation showed that common charge sensitive preamplifiers got saturated for few milliseconds. A working solution was found with the linear-logarithmic preamplifier MPR-16 produced by Mesytec [12]. This preamplifier features a nominal linear response up to 10 MeV and a logarithmic one above this energy. The energy deposited in the detector during the γ-flash was indeed very high, but due to the logarithmic response the recovery was faster, even though for some time during the recovery the preamplifier was in an intermediate state. It has to be remarked that the data acquisition system at n_TOF operates in waveform mode, and that the extraction of the useful information from the signals undergoes a digital filtering to account for possible baseline shifts. However, the closer one gets to the γ-flash, the more difficult (or even impossible) it becomes to distinguish signals of the $^{7}Be(n,\alpha)$ reaction from the tail of the γ-flash. In order to provide a quantitative assessment of Q3, a two-dimensional scatter plot was prepared showing the number of counts as a function of the neutron energy and of the energy measured by the back detector of the sandwich (Figure 7). The triton region is easily spotted as indicated by the dashed line. The transition from the linear to the logarithmic range of the preamplifier occurs between 5 and 10 keV (corresponding to flight times of 15 to 20 μs) as indicated by the down-bending of the triton locus. After selecting triton events inside the dotted line in Figure 7, a check was done trying to reproduce the cross section of the reaction (1) by normalizing to the known neutron flux in EAR2.

Figure 8 shows that the standard cross section [13] is perfectly reproduced up to few keV, whereas it starts to be seriously underestimated at energies above 10 keV.

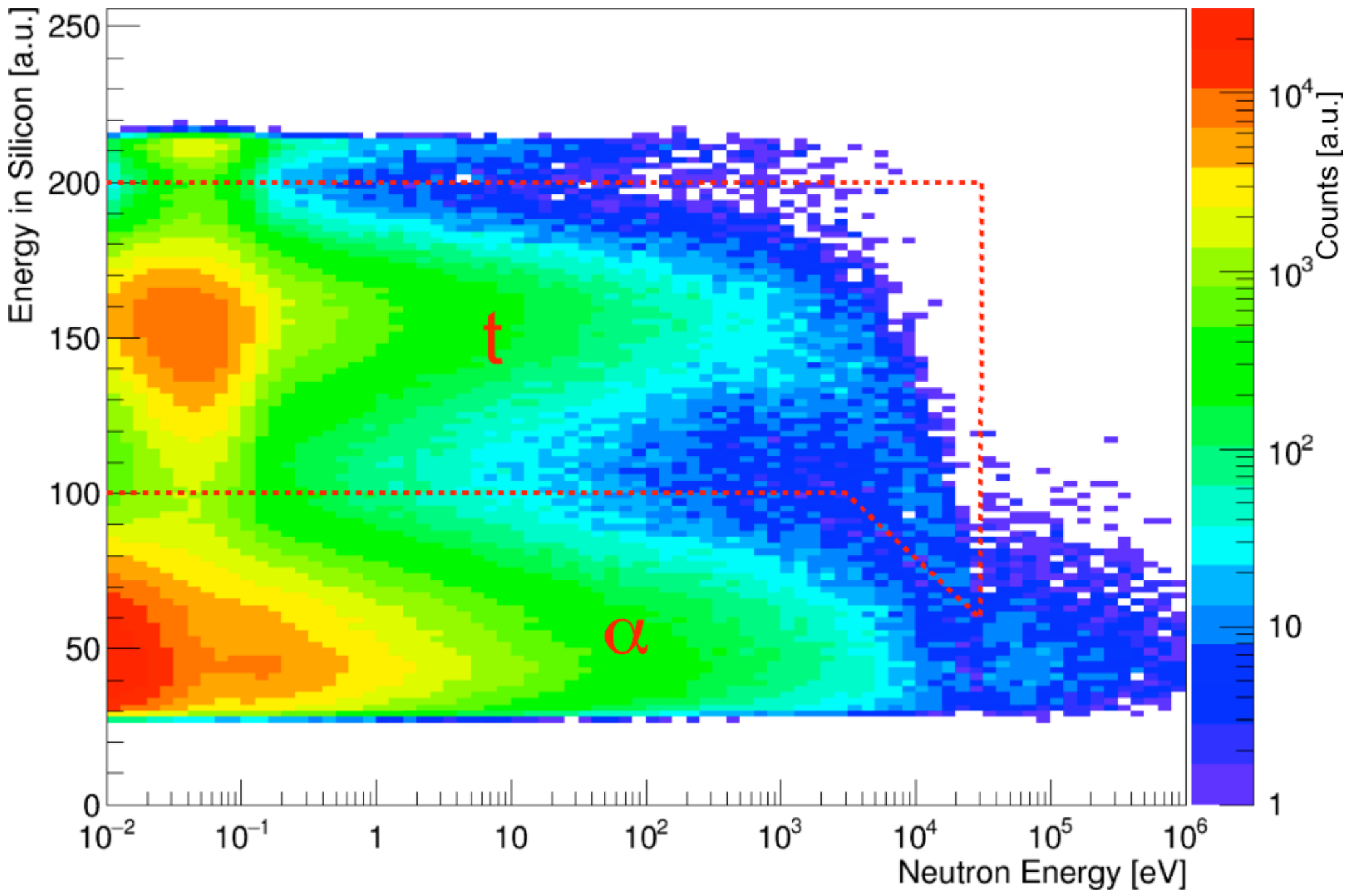

Figure 7. Distribution of counts as a function of the neutron energy and of the deposited energy in the back detector of the Silicon-$^{6}$LiF-Silicon sandwich. The triton region, enclosed by the dotted line, is clearly visible. See the text for further details.

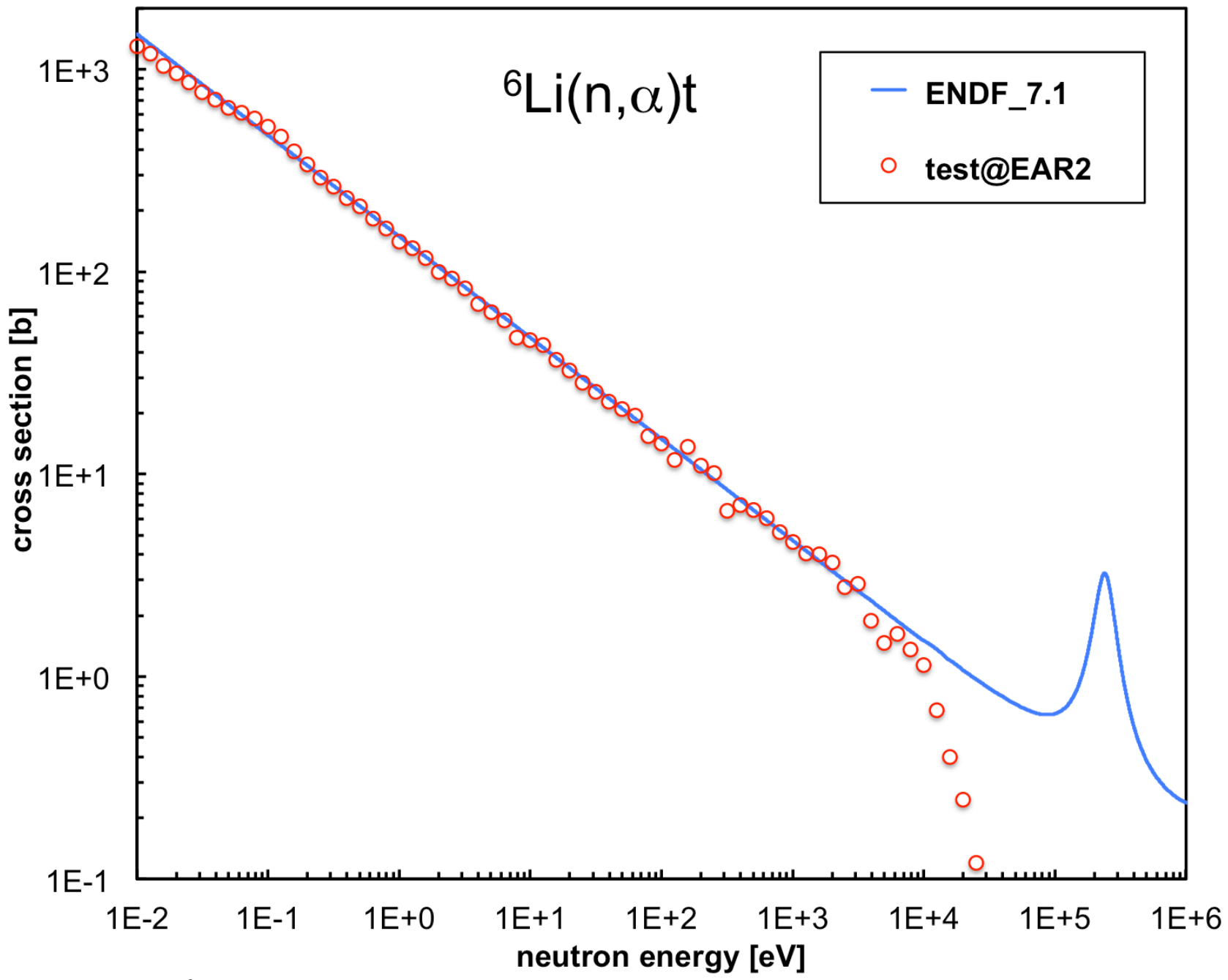

Figure 8. Reproduction of the $^{6}$Li(n,$\alpha$)t cross-section. At high neutron energy the detector signals occur very close to the blinding $\gamma$-flash, resulting in an increasing loss of useful signals. See the text for further details.

### *3.2 Test with a low-activity $^{7}$Be sample*

In response to question Q4 several tests have been performed at the target laboratory of PSI and in a hot room at LNS. To determine the counting rate from the $^{7}$Be $\gamma$-rays in the Silicon detectors, a vial containing a low activity $^{7}$Be solution (40.3 MBq, produced at PSI) was placed close to two Silicon detectors, namely PIPS_450 (Canberra PD450-19-100, 450 mm$^2$ active area, 100 µm thickness) and PIPS_300 (Canberra PD300-19-100, 300 mm$^2$ active area, 100 µm thickness). After calibrating the detectors with a $^{239}$Pu-$^{241}$Am-$^{244}$Cm alpha source, the two energy spectra shown in Figure 9 were acquired.

On top of the decreasing distribution, one can observe in both spectra the bump corresponding to the 478 keV gamma full energy peak from the $^{7}$Be decay. The observed overall counting rates were 1.5 and 1.3 kcps, respectively. The geometrical efficiency of PIPS_450 was ≈ 14%, and taking into account the 10% branching ratio the $\gamma$-ray detection efficiency was estimated to be of the order of ≈ $10^{-3}$. By upscaling this number to the final detector configuration one could reasonably expect a counting rate of the order of about 1÷10 Mcps. Obviously, the chosen electronic chain (preamplifier and amplifier) was not able to resolve single signals at such a counting rate, as the average time between $\gamma$-events would be about 0.1÷1 µs, not compatible with the ordinary front-end electronics for Silicon detectors. Two possible scenarios were unfolding: either the preamplifier would be totally paralyzed or it would produce a wide noise band due to the superposition of a huge number of incoming low energy signals. In order to investigate this issue, and to provide a realistic answer to question Q4, a test with a high intensity $\gamma$-source was needed.

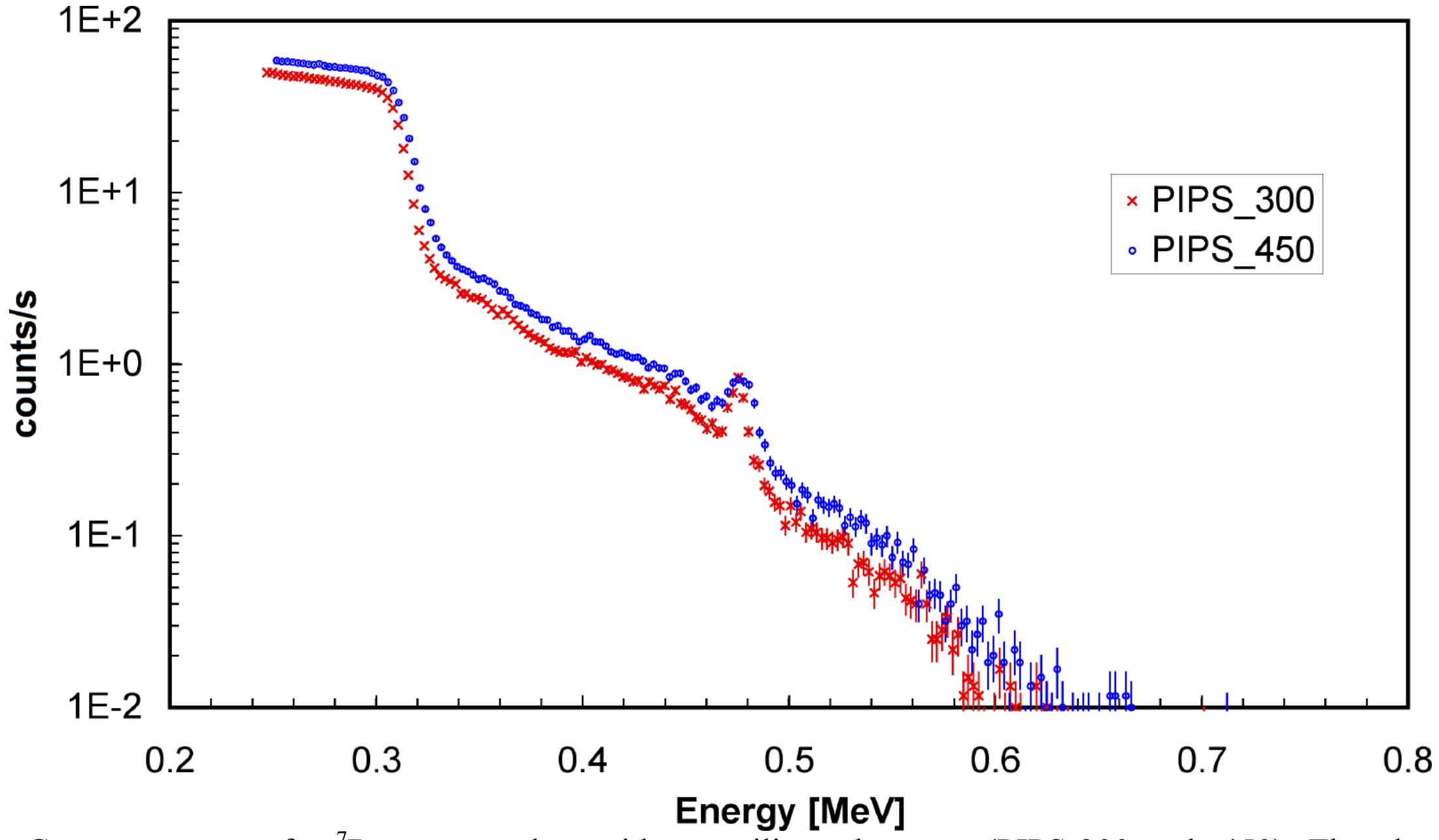

Figure 9. Gamma spectra of a $^{7}Be$ source taken with two silicon detectors (PIPS-300 and -450). The photopeak at 478 keV is visible, the total counting rates were respectively 1.3 and 1.5 kcps, respectively.

### *3.3 Test with a high intensity γ-source*

This final test was performed with a strong $^{137}Cs$ γ-source (39 GBq, 662 keV) available at LNS. The source is contained in a cylindrical shield and can be extracted by means of an electropneumatic actuator. in the setup of Figure 10 the detector position was tuned by means of a reference geiger counter such that about $10^{9}$ γ-rays per second were impinging on the detector. This should roughly correspond to 1÷10 Mcps expected in each Silicon detector from the $^{7}Be$ samples during the real measurement.

Again, the ordinary charge sensitive preamplifiers were immediately ruled out, as they were totally unresponsive due to the overwhelming event rate, whereas the Mesytec lin-log MPR-16 preamplifier proved capable of withstanding the counting rate even though showing a considerable noise. Under these background conditions the resolution was measured with the signals from an $^{241}Am$ alpha source, placed (in air) on top of the Silicon detector, with and without the high intensity $^{137}Cs$ γ-ray source, as shown in Figure 11. The presence of the huge γ-background did not prevent the discrimination of the ≈ 5 MeV alpha particles, only introducing a broadening of the peak and worsening the resolution by ≈ 200 keV. In the real experimental environment this effect is less relevant, due to the lower energy of the γ-rays (478 keV vs 662 keV) and to the higher energy of the alpha particles (8 MeV vs 5 MeV). A threshold around 2÷3 MeV was therefore considered sufficient for discriminating the expected γ-ray background.

As for the radiation damage, no appreciable effect due to γ-rays was observed on a time scale of one day, and the reverse current, which jumped from 20 nA to 40 nA with the $^{137}Cs$ source, went back to the initial value after removing the γ-source. It is well known, however, that the damage from neutrons is much more relevant than the one from γ-rays.

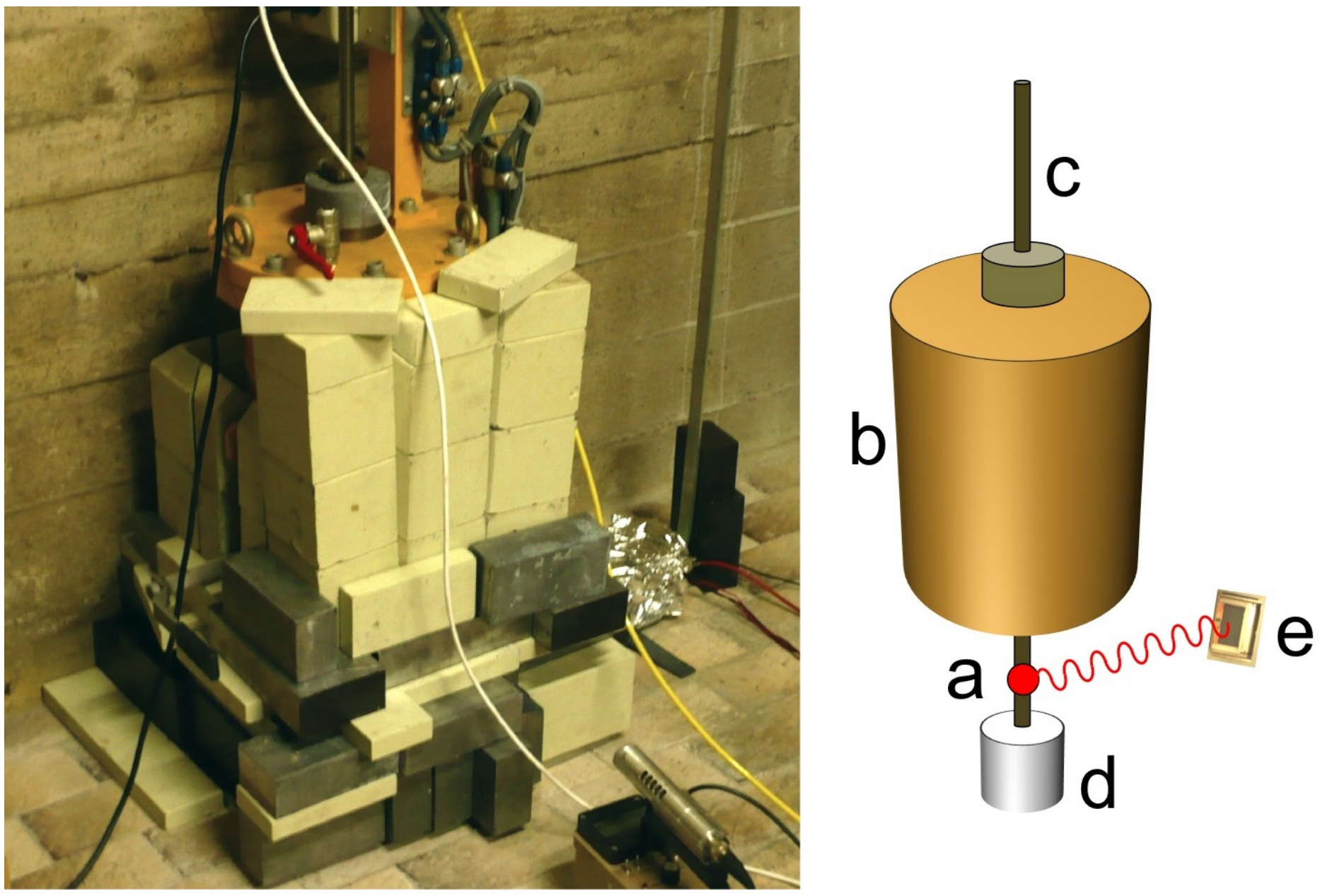

Figure 10. The test setup for the high intensity γ-irradiation with a 39 GBq $^{137}$Cs source (left) and the schematic sketch (right). (a) source; (b) shielding case; (c) electropneumatic actuator; (d) shielding plug; (e) Silicon detector.

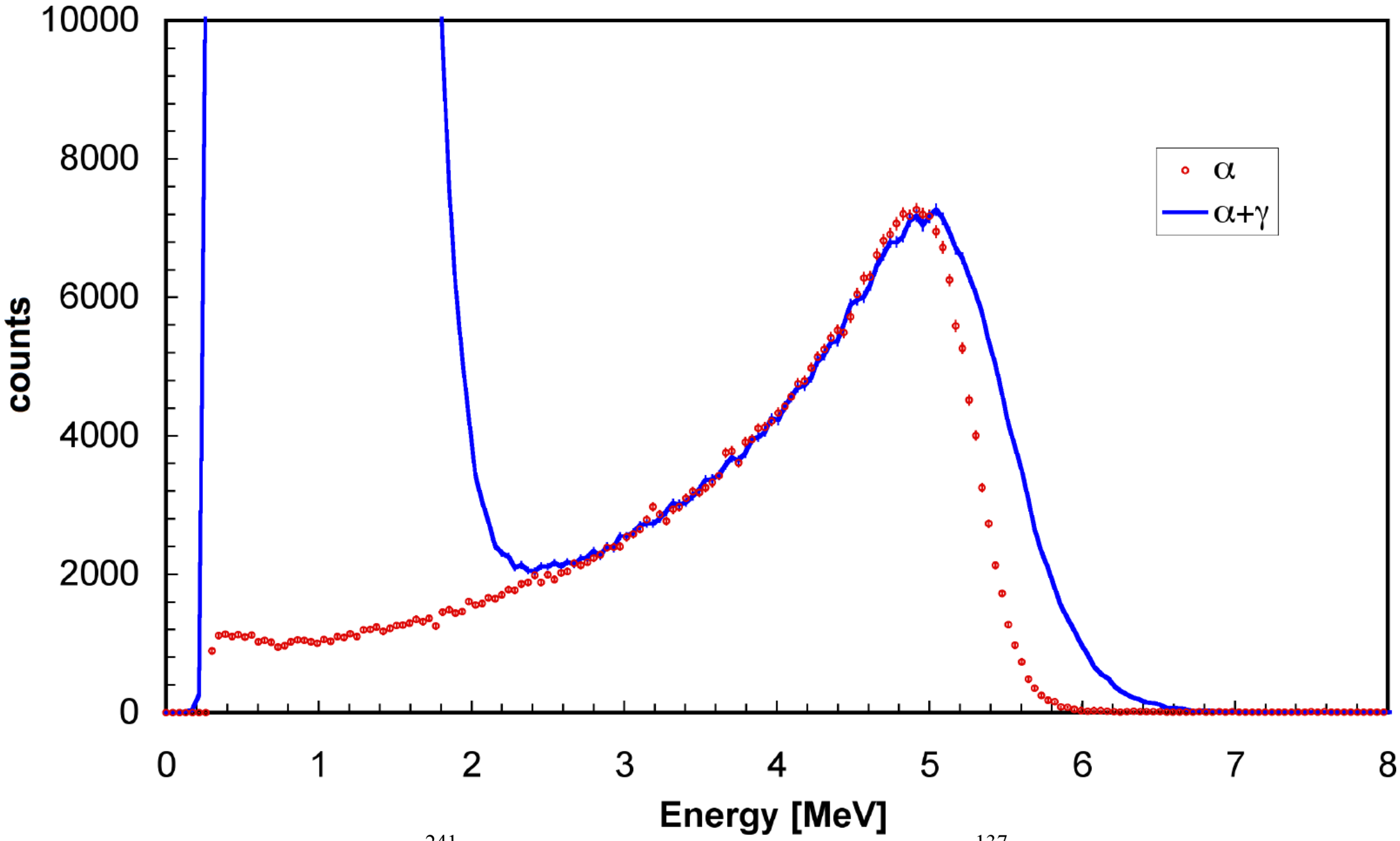

Figure 11. Alpha spectra taken with an $^{241}$Am source, with and without the high $^{137}$Cs γ-ray background produced in the setup of Figure 10.

## 4 Mechanical mounting and radioprotection issues

### *4.1 The chamber and the mechanical support*

The mechanical structure of the detector sandwiches used for the measurement of the $^{7}$Be(n,α) reaction at n_TOF was entirely made of carbon fibre, supported by four short threaded rods screwed to the base flange of the chamber (Figure 1). Each sandwich consists of seven U-shaped layers, of alternating width, thus providing slots for the two Silicon detectors and the target frame (Figure 12).

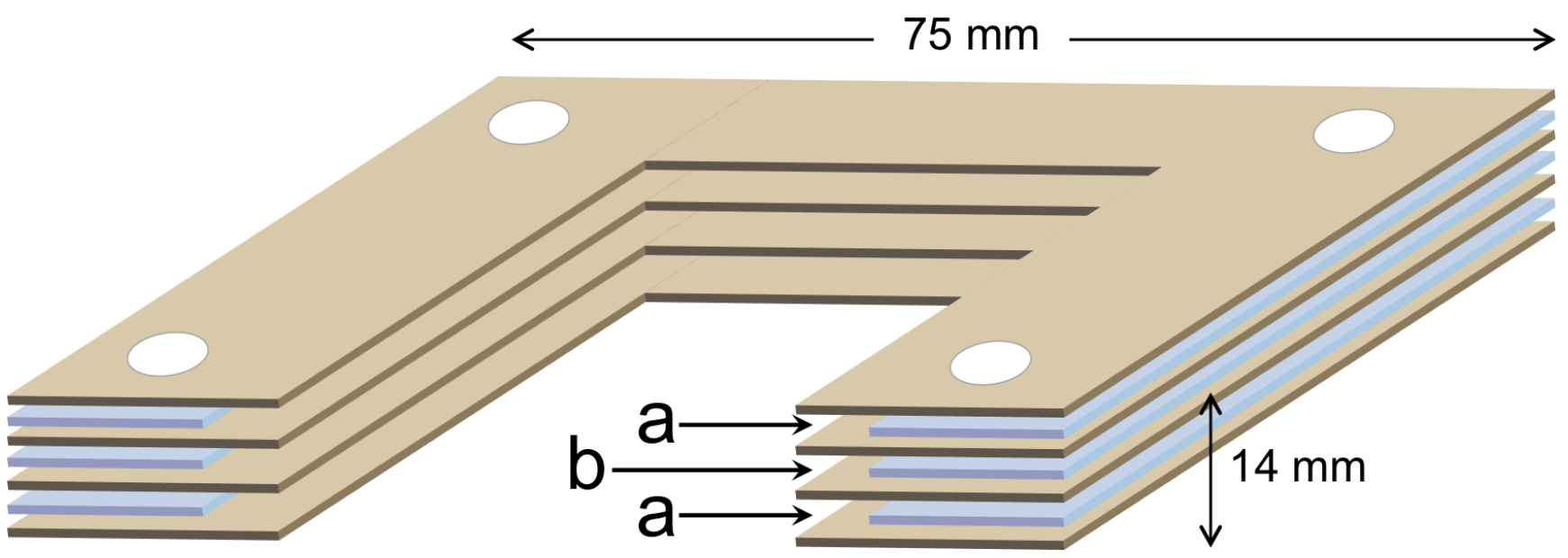

Figure 12. Sketch (not to scale) of the carbon fibre sandwich structure, composed of seven U-shaped layers of alternating width, thus providing slots for the two Silicon detectors (a) and the target frame (b).

The design was such as to allow for the initial insertion and assembling of the four Silicon detectors; then, after fixing the whole setup inside the chamber, the highly radioactive targets had to be inserted in a hot cell by means of a remote manipulator. A final flap, also in carbon fibre, was eventually latched via the manipulator to fix the targets in place (on the left in Figure 1). Due to the highly radioactive targets, the chamber containing the two sandwiches was heavily shielded with Lead, and had to be closed before any direct human intervention. The Lead shield was 1 cm thick, following a series of FLUKA simulations ([14],[15],[16]), as a tradeoff between outside dose rate, reasonable operating distance and time near the chamber, and its weight. The expected dose rate at 30 cm distance from a 50 GBq source of $^7$Be was evaluated to be below 1 mSv/h.

The chamber with its surrounding shield, sketched in Figure 13, was equipped with a pneumatic actuator in order to keep it open during the insertion of the targets. Figure 14 and Figure 15 show the chamber during setup and when ready for transport to the n_TOF beam line.

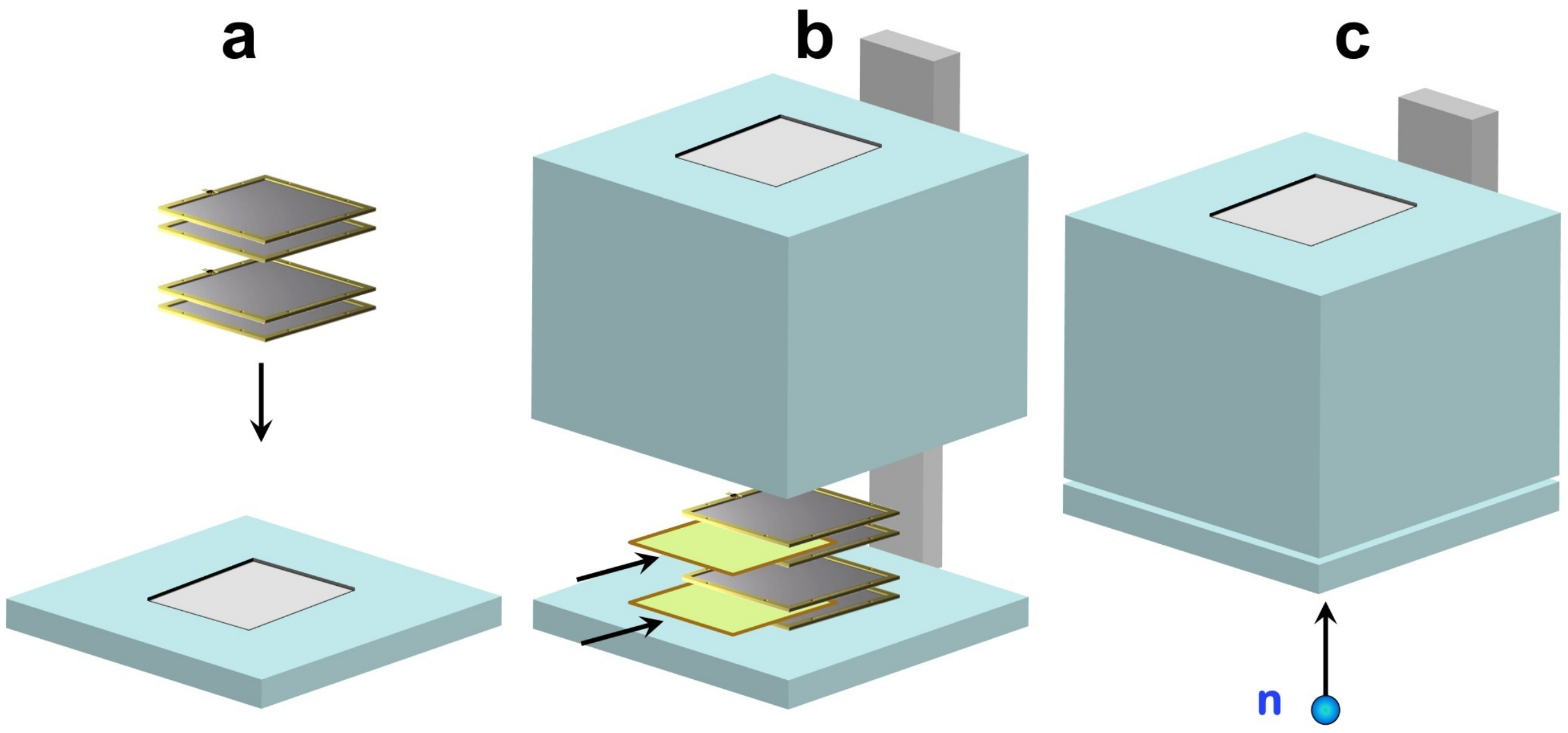

Figure 13. Assembling procedure of the reaction chamber. (a) The two detector sandwiches are assembled onto the chamber base, facing the thin beam entrance window. (b) The chamber body with the Lead shield is installed and kept open by means of a pneumatic actuator until the two radioactive $^7$Be targets are inserted amid the sandwiches by means of a remotely operated manipulator. (c) The chamber is closed, ready for transport and installation at the vertical n_TOF beam line.

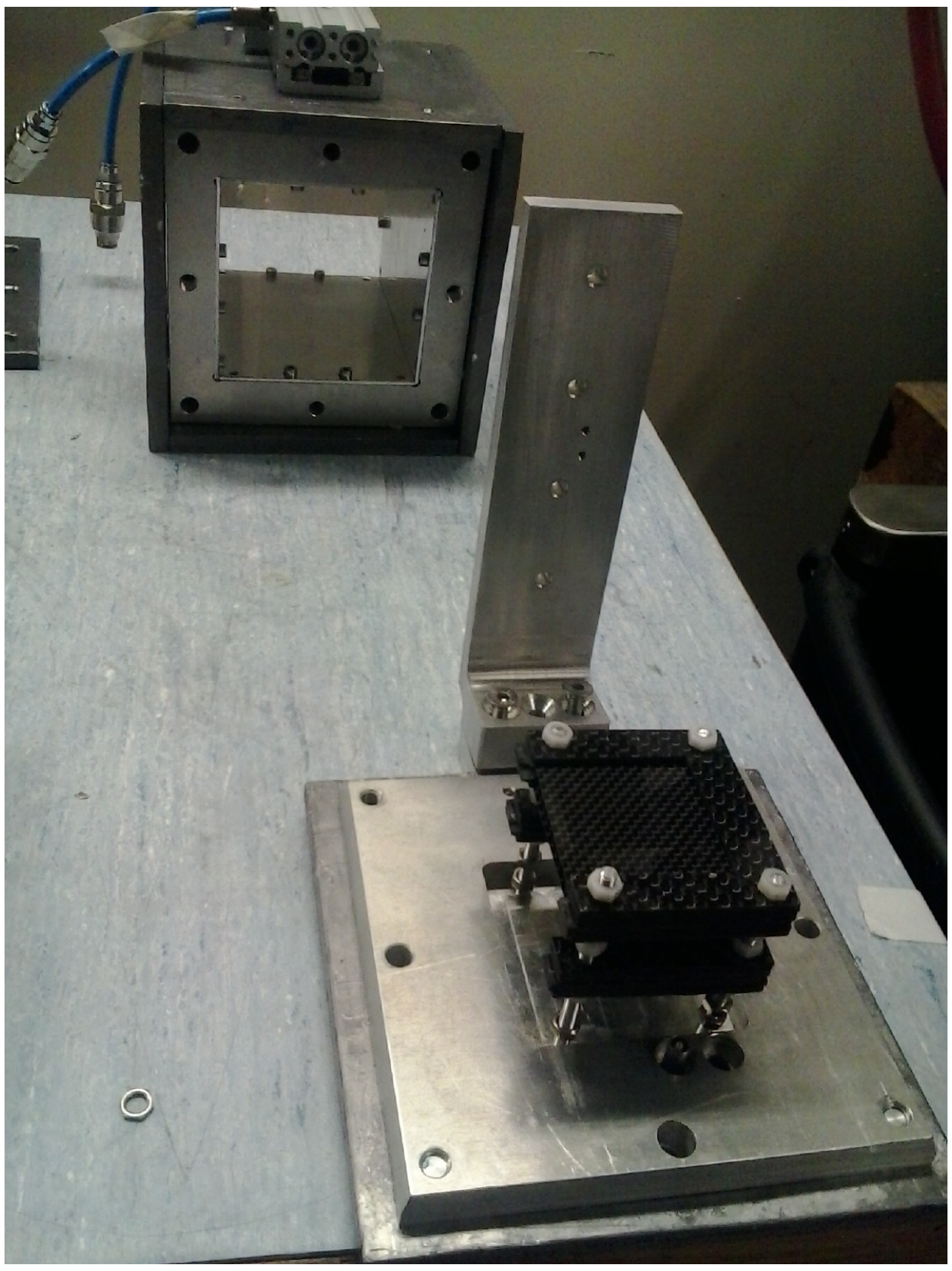

Figure 14. Background: the reaction chamber with the external Lead shielding. Foreground: the carbon fibre structure to host the two detector sandwiches assembled onto the chamber base.

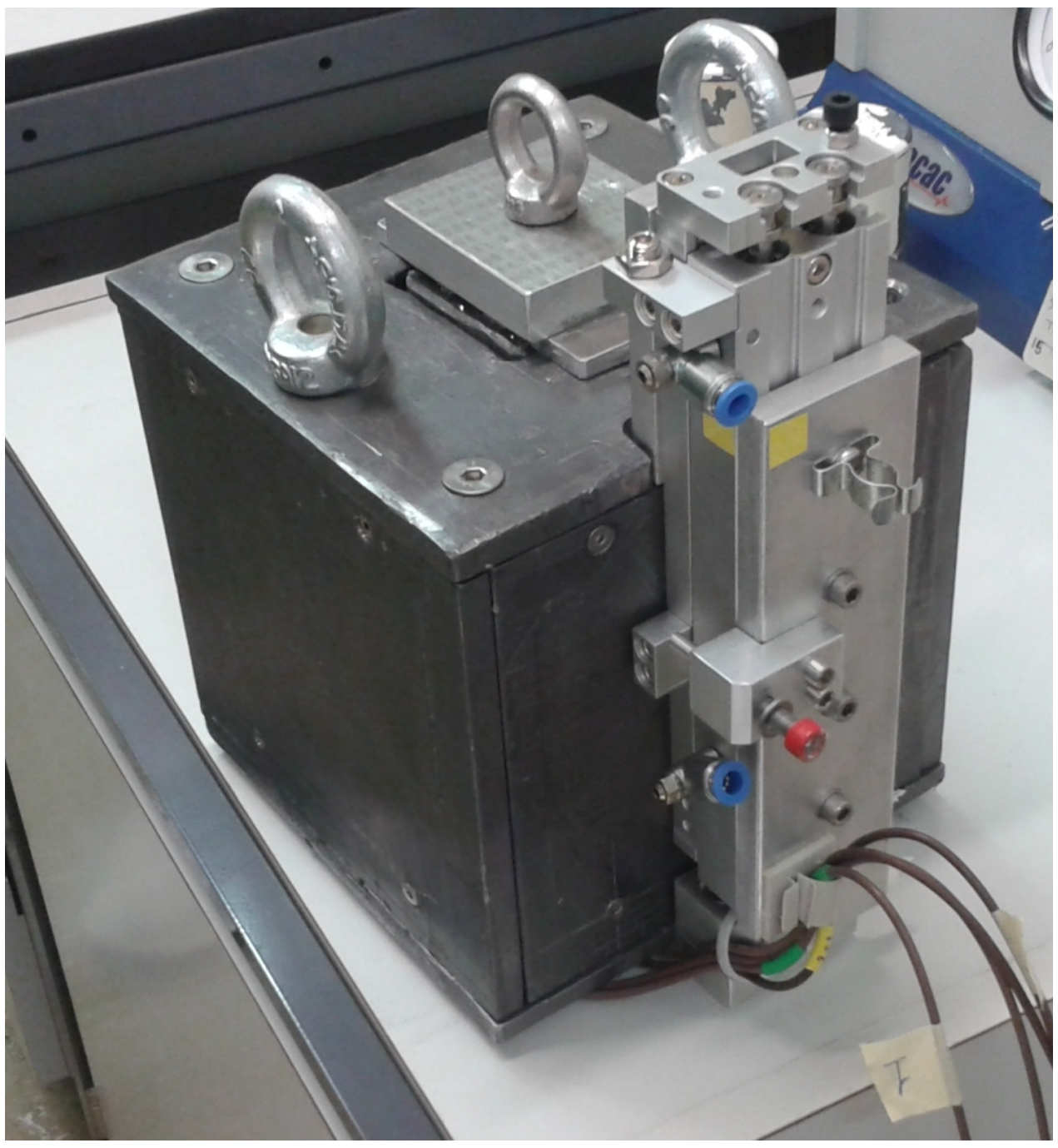

Figure 15. The reaction chamber after closure.

### 4.2 The $^7$Be target

A major challenge in the measurement of the $^7$Be(n,α)α cross-section was the availability of the $^7$Be in sufficient quantity. To achieve good statistical accuracy, a target containing a few micrograms of $^7$Be was needed, whereas in the only existing previous measurement the target mass was much less than a microgram. At PSI it is possible to extract and separate in one batch a relatively large amount of $^7$Be (up to several hundred GBq) from the cooling water of the SINQ spallation source [17].

Before proceeding with the real target, the preparation was tested by producing a dummy target of stable Be with a trace amount of $^7$Be. After purification from water and organic compounds, the Be was electrodeposited onto an Aluminum backing, with a loss of less than 0.05% as verified by means of γ-ray counting.

In order to evaluate possible systematic errors, it was decided to operate with a double setup, thus making use of two $^7$Be targets in the neutron beam, each one with its own set of detectors, to exploit the redundancy mentioned in section 2. Therefore, the final targets were produced with two different procedures. The first one was electrodeposited onto a 5 μm Aluminum foil and had a nominal activity of 18 GBq; the second one, with a nominal activity of 17 GBq, was deposited by droplet vaporization onto a 0.6 μm thick polyethylene foil glued onto a carbon fibre frame. The high $^7$Be activity required that all operations, including target production, were performed inside a hot-cell. A detailed description of the target production will be given in a separate paper [18].

## 5 The experiment

Following the target production at PSI, the chamber was transported to CERN for the experiment. The transport was organized using a certified container with an additional Lead shielding, to reduce the external dose rate to below 1 μSv/h at contact with the parcel. Once in n_TOF-EAR2, which is a Class A Laboratory, the chamber was removed from the container and installed on the neutron beam line, by means of a suitably prepared support structure, with minimal human intervention (estimated duration less than a minute). The cable connections to the prearranged connectors were done in few seconds, thus minimizing the absorbed dose in full agreement with the safety regulations.

The experiment was successfully performed from mid-August to the beginning of October 2015, with the detectors mostly behaving as expected. The reverse current was constantly monitored and, whenever needed, the voltage bias was raised in order to keep the detectors in full depletion mode compensating for the radiation damage and for the night/day temperature variations. A considerable number of coincidence events was detected as expected, and an example of two real signals is shown in Figure 16. The experimental data analysis is currently being finalized and the results will be published soon [8].

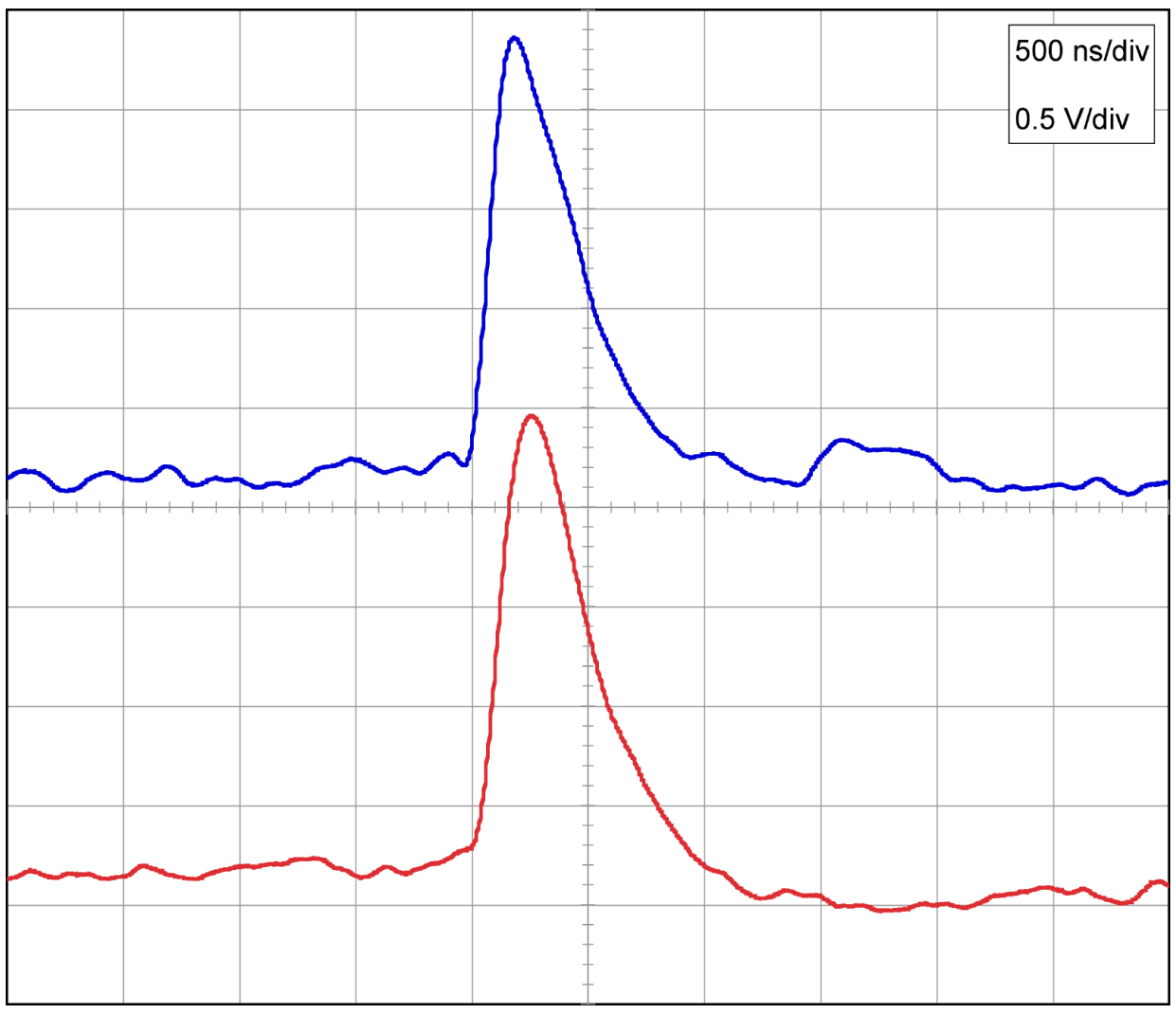

Figure 16. Snapshot of the two detector signals from an α–α coincidence event.

## 6 Conclusion

For the first time the $^7$Be(n,α)α reaction cross section could be measured in the neutron energy range from thermal up to several keV. Even though the energy range of astrophysical interest was not fully reached, this is a remarkable achievement, considering that the only previous information was at thermal energy published in 1963. The further extension of the operational energy range of the setup is being envisaged by developing a specialized electronics capable of recovering much faster after the γ-flash, likely allowing to reach the MeV neutron energy range. As the first measurement looks promising, an improved setup could soon be ready for a new measurement over a significantly extended energy range.

The detection techniques, the test procedures studied, proposed and developed for this work, the mechanical setup, the radioactive target production and handling, and the availability of the recently installed EAR2 neutron beam line, proved to be excellently working and could open the way to a new generation of experiments so far impossible to pursue.

## 7 Acknowledgments

The authors are grateful to Alexander Vögele (PSI) for managing the transport issues; Salvatore Russo, Giuseppe Passaro, Pietro Litrico and Angelo Seminara (INFN-LNS) for their invaluable support with the electronics and during the tests with the high activity $^{137}$Cs source.

This research was funded by the European Atomic Energy Communitys (Euratom) Seventh Framework Programme FP7/2007-2011 under the Project CHANDA (Grant No. 605203). We acknowledge the support by the Narodowe Centrum Nauki (NCN), under the grant UMO-2012/04/M/ST2/00700.